# Superlensing effect of an anisotropic metamaterial slab with near-zero dynamic mass

Xiaoming Zhou[1] and Gengkai Hu

*Key Laboratory of Dynamics and Control of Flight Vehicle, Ministry of Education,*

*School of Aerospace Engineering, Beijing Institute of Technology,*

*Beijing 100081, People's Republic of China*

A metamaterial slab of anisotropic mass with one diagonal component being infinity and the other being zero is demonstrated to behave as a superlens for acoustic imaging beyond the diffraction limit. The underlying mechanism for extraordinary transmission of evanescent waves is attributed to the zero mass effect. Microstructure design for such anisotropic lens is also presented. In contrast to the anisotropic superlens based on Fabry-Pérot resonant mechanism, the proposed lens operates without the limitation on lens thickness, thus more flexible in practical applications. Numerical modeling is performed to validate the proposed ideas.

[1] Electronic mail: zhxming@bit.edu.cn

Objects are often resolved from their scattering fields under wave excitations. Among the scattered fields, evanescent components with large spatial frequency carry the object's subwavelength features. Due to the decaying nature in conventional materials, evanescent waves are permanently lost in the image plane, resulting in the limited resolution of the imaging system. This is the well known diffraction limit,[1] which defines the spatial resolution not smaller than half of the operating wavelength. Since Pendry[2] proposed the optical perfect lens by negative refractive-index metamaterials, much efforts have been made to overcome the diffraction limit based on the metamaterial concept.[3,4] The fundamental mechanism for achieving the super-resolution effect is to enhance or preserve the amplitude of evanescent waves. Ambati et al showed that negative mass metamaterials could support surface resonant states.[5] Evanescent waves can be efficiently coupled to the surface state and their amplitudes are resonantly enhanced. A superlens with negative effective mass has been designed theoretically by acoustic metamaterials made of rubber-coated gold spheres in epoxy.[6] The coupling effect between the surface modes and evanescent waves will result in the nonuniform enhancement for evanescent waves of different spatial frequencies, therefore the created image may be distorted.

Acoustic metamaterials with anisotropic mass may provide an alternative to overcome the diffraction limit. Suppose a flat slab with a diagonal form of mass tensor $\tilde{\rho} = \text{diag}\left[\rho_\perp, \rho_\parallel\right]$. Its dispersion relation is of the following form

$$\frac{k_x^2}{\rho_\perp} + \frac{k_y^2}{\rho_\parallel} = \frac{\omega^2}{\kappa}, \tag{1}$$

where $k_x$ and $k_y$ are respectively the wave vectors perpendicular and parallel to the slab surface, $\kappa$ is the bulk modulus. Steel slabs placed periodically with constant slits has macroscopically an anisotropic mass with $\rho_\parallel = \infty$, and $\rho_\perp$ is around the mass density of the background material.[7,8] Infinite $\rho_\parallel$ arises from the extremely large impedance mismatch between the steel and the surrounding air. In such structural material, evanescent waves could propagate with the same wave vector $k_x = \omega\sqrt{\rho_\perp/\kappa}$ and the flat dispersion curve with respect to $k_y$ implies that evanescent waves with large parallel wave vector could be converted to propagating waves inside the anisotropic lens and transferred to the output side to form the super-resolution images. When the slab's thickness satisfies the Fabry-Pérot resonant condition, both propagating and evanescent waves achieve the complete transmission. Zhu et al designed delicate experiments, utilizing the hole-drilled brass block to obtain the spatial resolution $\lambda/50$ ($\lambda$ is the operating wavelength).[8] However the Fabry-Pérot resonant condition imposes the size limitation on the lens, i.e. the thickness should be an integer number of half-wavelengths. Therefore the anisotropic lens based on the Fabry-Pérot resonance is operative at the frequency dependent on the lens thickness.

In this letter, we will continue to examine the superlensing effect of anisotropic mass metamaterials. A new mechanism for efficient tunneling transmission of evanescent waves without above thickness limitation will be proposed based on the zero mass effect. Zero mass behavior has been observed in the discrete mass-spring system, where there are no phase variations of displacements between lattice units and high transmission can be achieved.[9] The material model of such anisotropic lens will also be presented and its super-resolution performance is validated by numerical modeling.

For a plane acoustic wave incident on a flat slab of anisotropic mass, the general expressions of transmission and reflection coefficients are given respectively by

$$t = \frac{4Z_x Z_{0x} e^{ik_x h}}{\left(Z_x + Z_{0x}\right)^2 - \left(Z_x - Z_{0x}\right)^2 e^{2ik_x h}}, \tag{2a}$$

$$r = \frac{\left(Z_x^2 - Z_{0x}^2\right)\sin(k_x h)}{\left(Z_x^2 + Z_{0x}^2\right)\sin(k_x h) + 2iZ_x Z_{0x} \cos(k_x h)}, \tag{2b}$$

where the wave impedances are $Z_{0x} = \omega \rho_0 / k_{0x}$ and $Z_x = \omega \rho_\perp / k_x$, $h$ is the slab's thickness, $k_{0x} = \sqrt{k_0^2 - k_y^2}$ and $k_0 = \omega\sqrt{\rho_0/\kappa_0}$, $\rho_0$ and $\kappa_0$ are the mass density and bulk modulus of the background material. From Eq. (2b), it is seen that there are two ways to eliminate the reflectance. One is based on the Fabry-Pérot resonant condition $k_x h = n\pi \ (n=1,2,...)$.[7,8] The other way, which this letter will focus on, comes from the impedance matching condition $Z_{0x} = Z_x$, from which we obtain the following relation

$$\left(1 - \frac{\rho_0^2}{\rho_\perp \rho_\parallel}\right) k_y^2 = \omega^2 \rho_0^2 \left(\frac{1}{\rho_0 \kappa_0} - \frac{1}{\rho_\perp \kappa}\right). \tag{3}$$

If the impedances are matched for arbitrary $k_y$, the following two relations should be satisfied

$$\rho_0^2 = \rho_\perp \rho_\parallel, \tag{4a}$$

$$\rho_0 \kappa_0 = \rho_\perp \kappa. \tag{4b}$$

As already indicated, $\rho_\parallel = \infty$ is necessary for achieving the flat dispersion curve, thus Eq. (4a) suggests $\rho_\perp \to 0$. Taking the limit $\rho_\perp \to 0$ and $\rho_\parallel \to \infty$ in Eq. (2), the transmission and reflection coefficients are rewritten as

$$t = \frac{1}{1 - \dfrac{i\pi h/(\alpha \lambda_0)}{\sqrt{1 - k_y^2/k_0^2}}}, \tag{5a}$$

$$r = t - 1, \tag{5b}$$

where $\alpha = \kappa/\kappa_0$ and $\lambda_0$ is the wavelength in the background material. In order that high transmission is achieved, it is necessary that $h/(\alpha \lambda_0) \ll 1$. One possibility for this condition being satisfied is to increase $\alpha$, and when $\alpha$ is sufficiently large, the second requirement (4b) for the impedance matching will be fulfilled, leading to the total transmission. The other choice is to decrease the relative thickness $h/\lambda_0$ of the slab. This process aims to make the lens as acoustically thin as possible and high transmission can be explained by the mass law, which states that the transmittance is inversely proportional to the material thickness. Similar mechanism has been found in electromagnetic waves and energy can tunnel through epsilon-near-zero subwavelength materials.[10] Above analyses demonstrate that a metamaterial slab of anisotropic mass $\rho_\perp \to 0$ and $\rho_\parallel \to \infty$ can enable efficient transmission of both propagating and evanescent waves as long as the condition $h/(\alpha \lambda_0) \ll 1$ is satisfied. Therefore the superlensing effect can be expected in such anisotropic material. The operating frequency of such lens will be dependent on the way zero mass is realized. In what follows, metamaterial realization of the lens is proposed and zero effective mass can be designed to be invariant to the lens thickness.

The designed lens consists of solid slabs with a periodic array of slits partially filled by elastic layers, as shown in Fig.1. The width of slit is $a$ and the lattice constant of the grating period is $d$. In each slit, elastic layers of the thickness $w$ are separated by cavities, forming a periodic array with lattice parameter *s*. The thickness of the lens is *h=Ns*. Consider the case where the periodicities *d* and *s* are much less than the operating wavelength, the dynamic property of the lens can be well characterized by effective medium theory. The solid slabs are all fixed, so that infinite effective mass of the lens in the *y* direction is strictly satisfied. In the *x* direction, clamped elastic layers inside the slits undergo resonances at their lowest eigenfrequency. It has been shown that the resonant vibration of clamped layers will result in effective mass of metamaterials that follows the Drude-form expression $\rho_\perp = \rho(1-\omega_0^2/\omega^2)$, where $\omega_0$ is the cutoff frequency.[11] Zero effective mass can be realized at $\omega_0$.

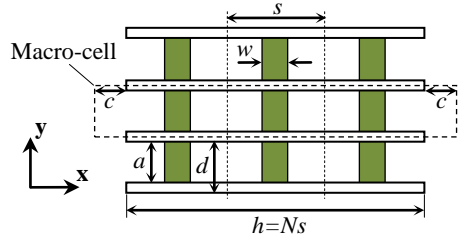

FIG. 1. (Color online) Schematic of the proposed lens made of fixed solid slabs with a periodic array of slits partially filled by elastic layers.

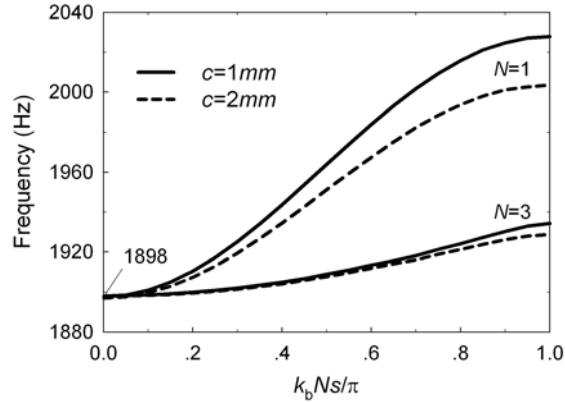

FIG. 2. The dispersion curves given by normalized Bloch wave vector versus frequency for the macro-cell unit of the slab lens with different thickness *h=Ns* in different separation *c*.

As an example, the geometric sizes of the lens are taken as *a*=4mm, *d*=5mm, *w*=2mm, and *s*=10mm. The elastic layer is the rubber[12] with Young's modulus 1MPa, Possion's ratio 0.49, and mass density 1100kg/m³. Mass density and sound velocity of the air surrounding and cavities are taken to be $\rho_0 = 1.25$ kg/m$^3$ and $c_0 = 343$ m/s. Numerical modeling is performed based on commercial software COMSOL Multiphysics. To determine the zero-mass frequency, the dispersion curve is computed for the macro-cell unit, as depicted in Fig.1, by use of Floquet-Bloch periodic conditions imposed on the left and right boundaries. Figure 2 shows the lowest branch of the dispersion curves for different lens thickness $h=Ns$ and separation $c$. According to general dispersion relation $k_b = \omega\sqrt{\rho_\perp / \kappa}$, zero effective mass exists at the frequency where the Bloch wave vector $k_b$ is zero. It can be found that the zero mass frequencies are identically 1898 Hz for different modeling parameters, because the cutoff frequency at which effective mass is zero is defined by the eigenfrequency of the clamped rubber layer. The identical zero-mass frequency for different period $N$ reveals the robustness of the designed lens in that the operating frequency will be invariant to the lens thickness. To verify the imaging effect, two monopole line sources operating at frequency 1898Hz (the corresponding wavelength is denoted by $\lambda_0$) separated by 20mm ($\lambda_0/9$) are placed in front of the lens at the distance 1mm and the image plane is taken 1mm behind the lens. Figure 3 shows the contour plots of pressure amplitude distributions at the image plane and frequencies ranging from 1897Hz to 1900Hz for the lens with the thickness $h=w$. From pressure distributions, two sources can be clearly resolved at zero-mass frequency, confirming the super-resolution imaging of the designed lens beyond the diffraction limit. Note that at 1898Hz, effective bulk modulus $\kappa$ is 0.18 MPa,[13] then $h/(\alpha\lambda_0)=0.045$. From Eq. (5a), there is a singular point $k_y = k_0$ in the transmission spectrum around which the transmission is nonuniform. It is worth to note that this distorted band can be greatly narrowed in an optimally designed metamaterial with $h/(\alpha\lambda_0)$ far less than unity. It can be observed in Fig. 3 that there is a finite bandwidth of operating frequencies. As the frequency increases away from 1898Hz, the images are gradually degraded because effective mass becomes a positively nonzero value, and the transmission amplitudes are lowered. The low edge of the band is limited by surface wave resonances excited by negative effective mass[5] in the way that bright spots are produced to make the image difficult to distinguish.

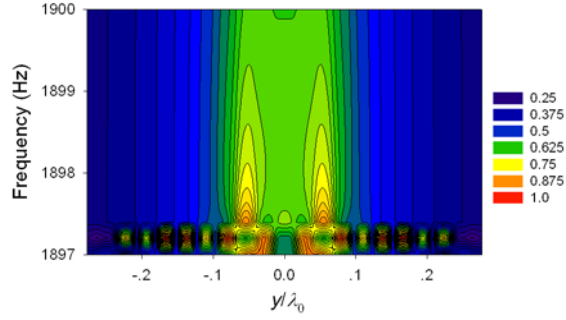

FIG. 3. (Color online) Contour plot of normalized pressure amplitude distributions taken 1mm behind the slab lens of thickness $h=w$ at different frequencies for two monopole line sources separated by $\lambda_0/9$ being placed in front of the lens at the distance 1mm.

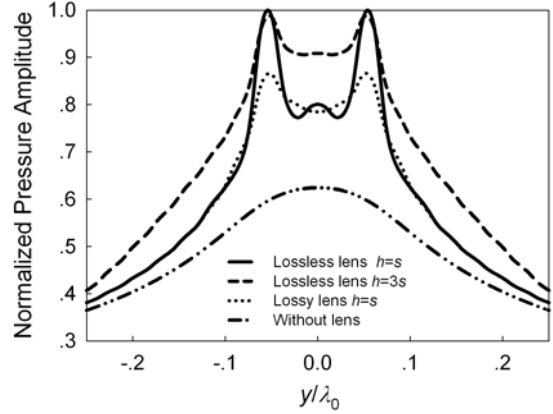

FIG. 4. Normalized pressure amplitudes in the image plane calculated at 1898Hz in four cases: imaging with the lossless superlens of thickness $h=w$ and $h=3w$, and with and without the lossy lens of thickness $h=w$.

At frequency 1898Hz, the normalized pressure amplitudes in the image plane by the slab lens of different thicknesses $h=w$ and $h=3w$ are shown in Fig. 4. It is seen that without the change of operating frequency the superlens with larger thickness can also offer enough contrast for the images. This performance breaks the thickness limitation existing in the anisotropic lens based on Fabry-Pérot resonant mechanism. Further, to study the influence of the material loss, the imaging property of a lossy lens with thickness $h=w$ is presented in Fig. 4, where the loss factor 2.0e-4 is added to the Young's modulus of the rubber. Effective mass density of such lossy lens is calculated to be $\rho_\perp=0.056i$.[13] For comparison, the result computed in the absence of the lens is also given.

It is found that the super-resolution imaging is still achieved and the material damping will degenerate the imaging quality. In this example, we do not optimize the microstructures for the low-loss macroscopic behavior. In practical realization, weak damping can be achieved by low-loss metals, such as aluminum or silver membranes, in place of rubber materials.

In conclusion, we have demonstrated the superlensing effect in anisotropic materials with zero mass density. In contrast to the anisotropic superlens based on the Fabry-Pérot resonance, the operating frequency is independent of the thickness in the proposed lens. In this study, the theory and microstructure design of the superlens is based on the two-dimensional problem, but can be easily extended to the three-dimensional case. Thus the experimental realization is very promising. Applications based on such anisotropic superlens can be anticipated in the areas of medical imaging and non-destructive evaluation.

This work was supported by the National Natural Science Foundation of China (10832002, 10702006) and the National Basic Research Program of China (2006CB601204).

effective bulk modulus $\kappa$ and effective mass density $\rho_\perp$ are computed by

$$\kappa = \iint_S (\sigma_{ii}/3) dS / \left( \int_{\text{right}} u dl - \int_{\text{left}} u dl \right) \text{ and } \rho_\perp = \left( \int_{\text{right}} p dl - \int_{\text{left}} p dl \right) / \left( \omega^2 \iint_S u dS \right),$$

where $\sigma_{ij}$ is the stress tensor, $u$ is the x-component of the displacement, $p$ is the pressure, the subscripts 'S', 'left', and 'right' represent respectively the integration over the surface, left and right boundaries of the unit.